# ApexGen: Simultaneous design of peptide binder sequence and structure for target proteins

Xiaoqiong Xia[1-4] and Cesar de la Fuente-Nunez[1-4,*]

[1]Machine Biology Group, Departments of Psychiatry and Microbiology, Institute for Biomedical Informatics, Institute for Translational Medicine and Therapeutics, Perelman School of Medicine, University of Pennsylvania, Philadelphia, Pennsylvania, United States of America.

[2]Departments of Bioengineering and Chemical and Biomolecular Engineering, School of Engineering and Applied Science, University of Pennsylvania, Philadelphia, Pennsylvania, United States of America.

[3]Department of Chemistry, School of Arts and Sciences, University of Pennsylvania, Philadelphia, Pennsylvania, United States of America.

[4]Penn Institute for Computational Science, University of Pennsylvania, Philadelphia, Pennsylvania, United States of America.

*Correspondence:

Cesar de la Fuente-Nunez (cfuente@upenn.edu)

## Abstract

Peptide-based drugs can bind to protein interaction sites that small molecules often cannot, and are easier to produce than large protein drugs. However, designing effective peptide binders is difficult. A typical peptide has an enormous number of possible sequences, and only a few of these will fold into the right 3D shape to match a given protein target. Existing computational methods either generate many candidate sequences without considering how they will fold, or build peptide backbones and then find suitable sequences afterward. Here we introduce ApexGen, a new AI-based framework that simultaneously designs a peptide's amino-acid sequence and its three-dimensional structure to fit a given protein target. For each target, ApexGen produces a full all-atom peptide model in a small number of deterministic integration steps. In tests on hundreds of protein targets, the peptides designed by ApexGen fit tightly onto their target surfaces and cover nearly the entire binding site. These peptides have shapes similar to those found in natural protein–peptide complexes, and they show strong predicted binding affinity in computational experiments. Because ApexGen couples sequence and structure design at every step of Euler integration within a flow-matching sampler, it is much faster and more efficient than prior approaches. This unified method could greatly accelerate the discovery of new peptide-based therapeutics.

## Introduction

Peptide therapeutics address a critical gap in druggable target space by engaging protein–protein interfaces that remain inaccessible to small molecules while avoiding the manufacturing complexity of larger biologics[1–3]. However, designing potent peptides is intrinsically challenging. A peptide of length $L$ spans approximately $\sim 20^L$ possible sequences distributed across a vast conformational continuum, yet therapeutic efficacy typically demands precise three-dimensional complementarity to a specific receptor surface[4]. Thus, discovery requires joint optimization of sequence and backbone geometry under the constraints imposed by the target receptor[5].

Existing discovery strategies interrogate only a fraction of this space. Library-based selections (e.g. phage display)[6] and physics-based approaches (molecular docking and dynamics)[7] have yielded successes, but they require extensive target-specific validation and computational tuning[8]. Deep learning can partially alleviate scaling bottlenecks. For example, sequence-only generative models (recurrent neural networks, variational autoencoders, Transformers) can propose vast candidate libraries[9–13], but they remain fundamentally structure-agnostic and thus require costly retrospective structure prediction or docking[14,15]. Structure-aware diffusion models (e.g. RFdiffusion)[16,17] can design protein backbones and interfaces, but they rely on stochastic multi-step denoising and complex equivariant architectures, which slow sampling and often decouple sequence design from backbone generation (e.g., via post-hoc ProteinMPNN application)[18]. Recent approaches like PepFlow and ppflow[19,20] advance simultaneous sequence-structure generation, bu they typically require post-hoc geometry correction to satisfy stereochemical constraints.

Flow matching offers a deterministic alternative to stochastic diffusion. It learns time-dependent vector fields whose ordinary differential equation (ODE) trajectories transport simple prior distributions to empirical data distributions[21,22]. This framework enables simulation-free training and fast, deterministic sampling via coarse ODE integration. Riemannian extensions naturally operate on molecular manifolds including tori (for dihedral angles) and SO(3) (for rigid-body rotations)[23], making it well-suited to peptide design, where data reside on the product manifold spanning backbone torsions, rigid frames, and discrete sequence identities[24].

Here we introduce ApexGen, a receptor-conditioned, manifold-aware conditional flow-matching framework for all-atom peptide sequence-structure co-design. ApexGen learns four coupled vector fields that govern: (i) backbone dihedral angles ($\varphi, \psi, \omega$) on a high-dimensional torus; (ii) per-residue rigid-body frames in SE(3); (iii) amino-acid identities on the probability simplex; and (iv) side-chain dihedral angles on a torus, with sequence-dependent masking to ensure physical plausibility. We enforce geometric consistency by reconstructing all-atom coordinates at each training step and applying stereochemical violation losses inspired by OpenFold[25]. Benchmarking ApexGen on 613 whole-protein and 165 binding-pocket receptor targets, we demonstrate deterministic generation with higher structural fidelity (TM-score 0.99 versus 0.33) and interface complementarity (binding-site ratio 0.94) compared to other multi-manifold baselines: PepFlow. Designed peptides recapitulate native backbone Ramachandran basin structure, exhibit chemically interpretable amino-acid preferences, and show favorable predicted binding energetics, providing a computationally efficient platform for structure-based peptide therapeutic discovery.

## Methods

**Problem setting and notation**

We formulate peptide design as learning a trajectory on a product manifold spanning torsional, rigid-body, and discrete sequence spaces. Consider a peptide of length $L$ residues. We represent its state as $x = (\alpha_{bb}, g, p, \alpha_{sc})$, where $\alpha_{bb} = (\varphi, \psi, \omega)$ denotes backbone dihedral angles residing on a 3L-dimensional torus $\mathbb{T}^{3L}$ ($\varphi$: N–C$\alpha$–C bond rotation, $\psi$: C$\alpha$–C–N bond rotation, $\omega$: peptide-bond torsion). $g = \{R_i, t_i\}^L$ comprises per-residue rigid-body frames with rotation matrices $R_i \in$ SO(3) and translation vectors $t_i \in \mathbb{R}^3$, collectively forming elements of SE(3)$^L$. $p = p_i^L, p \in \Delta^{20}$ represents amino-acid probability distributions over 20 canonical residue types (the 20-dimensional simplex); and $\alpha_{sc} = (\chi_1, \chi_2, \chi_3, \chi_4)$ encodes side-chain dihedral angles on $\mathbb{T}^{4L}$, with undefined $\chi$ angles masked according to residue identity. The receptor context $c$ provides sequence and structural features extracted from the target protein. Mathematically, the state manifold is

$$\mathcal{X} = \mathbb{T}^{3L} \times \text{SE}(3)^L \times \Delta^{20L} \times \mathbb{T}^{\leq 4L}.$$

This multimodal representation enables simultaneous optimization of backbone geometry, frame placement, residue identity, and side-chain conformation under receptor-imposed constraints.

**Conditional flow matching on product manifolds**

Conditional flow matching regresses analytic velocities of geometry-consistent interpolants between prior and data states. Given an initial state $x_0 \sim q_0$ (a tractable prior) and a target $x_1 \sim p_{data}$ (empirical data), we define geometry-consistent interpolations $\gamma(t; x_0, x_1)$ for each modality and regress the analytic velocities using conditional flow matching[23,24]. The training objective is

$$\mathcal{L}_{\text{CFM}} = \mathbb{E}_{(x_1,c)\sim\mathcal{D}} \, \mathbb{E}_{x_0\sim q(\cdot|x_1)} \, \mathbb{E}_{t\sim\mathcal{U}(0,1)} \, \| v_\theta(x_t, t, c) - \partial_t \gamma(t; x_0, x_1) \|^2_{\mathcal{T}_{x_t}\mathcal{X}},$$

where $x_t = \gamma(t; x_0, x_1)$ and the norm is computed in the tangent space $\mathcal{T}_{x_t}\mathcal{X}$. Unlike stochastic diffusion, conditional flow matching eliminates noise schedules and enables deterministic ODE integration during sampling. Stochastic interpolants unify flows and diffusions, motivating our interpolation choices[22]. On curved spaces, we employ Riemannian flow matching for geometric consistency[23].

**Modality-specific paths**

Each modality follows geodesics or linear interpolants tailored to its geometric structure. Translations use linear interpolation with constant velocity: $t(t) = (1-t)t_0 + tt_1$, yielding velocity $v_t^* = t_1 - t_0$. Rotations follow shortest-arc geodesics in SO(3) via the matrix exponential and logarithm: $R(t) = \exp(t\Omega)R_0$, where $\Omega = \log(R_1 R_0^T)$, with tangent-space velocity $v_R^* = \Omega \in \mathfrak{so}(3)$. SO(3) geodesics preserve rotation distance and avoid gimbal-lock singularities inherent to Euler-angle representations. Backbone and side-chain angles evolve on the torus using wrapped linear interpolation: $a(t) = \text{wrap}((1-t)a_0 + ta_1)$, with velocity $v_a^* = \text{wrap}(a_1 - a_0)$. Wrapping ensures continuity across the $\pm\pi$ boundary, preventing artificial energy barriers. For residue identities, we interpolate in logit space: $\ell(t) = (1-t)\ell_0 + t\ell_1$, then project to the simplex via $p(t) = softmax(\ell(t))$, with velocity $v_\ell^* = \ell_1 - \ell_0$. This parameterization maintains probabilistic constraints throughout the trajectory.

**Network architecture**

**Overview.** ApexGen comprises three components: a receptor encoder that builds residue-level node and edge embeddings for the receptor–peptide complex with masking to prevent peptide ground-truth leakage; a sinusoidal time encoder that broadcasts temporal context; and a stacked Invariant Point Attention (IPA)-Transformer denoiser that conditions peptide updates on receptor features (Figure 1). The denoiser trunk comprises multiple IPA blocks followed by per-residue readouts yielding four modality-specific predictions: rotation, translation, sequence logits, and dihedral angles. Information flows from receptor features through IPA blocks to modality-specific output heads that predict velocities on their respective manifolds.

**Receptor encoder.** The encoder aggregates atomic geometry, sequence identity, and pairwise relationships into residue-level embeddings. For each receptor residue, we construct node embeddings by concatenating: (i) learned amino-acid identity embeddings; (ii) local atomic geometry (backbone and $C_\beta$ coordinates) expressed in a residue-local frame derived from Cα-C-N atoms; and (iii) backbone dihedral angles ($\varphi, \psi, \omega$) encoded with multi-frequency sinusoidal maps ($\sin(k\varphi), \cos(k\varphi)$ for $k = 1, \ldots, K$). These features are projected by a multi-layer perceptron into a unified node representation. Masking prevents peptide ground-truth information from entering receptor embeddings during training.

For each residue pair, we construct edge embeddings by fusing: (i) pairwise amino-acid type embeddings; (ii) clipped relative sequence offset (clamped to ±32); (iii) inter-residue distance features from heavy-atom pairs expanded with a Gaussian radial basis function; and (iv) pairwise dihedral angles between residue frames encoded with the same multi-frequency map. Edge masks prevent peptide-peptide interactions from leaking ground-truth geometry.

**Time encoder.** We embed the scalar time coordinate $t \in [0,1]$ using sinusoidal positional encoding: $[\sin(2^k \pi t), \cos(2^k \pi t)]$ for $k = 0, \ldots, K_{time}$. The encoded time features are broadcast to all residues and concatenated with node embeddings before each IPA block, providing temporal conditioning throughout the network.

**IPA-Transformer denoiser: conditioning on receptor node/edge features.** Each IPA block updates peptide residue representations while jointly attending to peptide and receptor nodes. The invariant point attention mechanism forms queries from the current peptide state and keys/values from the union of peptide and receptor states[25]. We inject receptor pairwise information by adding a linear projection of the edge embedding $e_{ij}$ to the attention logits before softmax normalization. Structural masks $m_{ij}$ prevent attention to spatially distant or temporally inappropriate residues.

Point-valued features (coordinate triplets in local frames) are aggregated via weighted summation, then transformed back to global coordinates using the current frame estimate. This yields an SE(3)-equivariant update of the peptide representation. Each block applies a feed-forward node-transition MLP followed by a rigid-frame update that parameterizes an SE(3) velocity per residue: rotation velocity $\Omega_i \in \mathfrak{so}(3)$ and translation velocity $t_i \in \mathbb{R}^3$. Frame updates apply only to peptide residues; receptor frames remain fixed and serve purely as conditioning context.

**Output heads.** From the final single-representation vector of each peptide residue, we emit four modality-specific predictions. The rotation head outputs a skew-symmetric matrix $\Omega_i \in \mathfrak{so}(3)$ that parameterizes the SO(3) velocity for rigid-frame rotation. The translation head outputs a 3D vector $t_i \in \mathbb{R}^3$ representing the frame origin velocity. The sequence head produces 21-way logits $\ell_i$ (20 amino acids plus a mask token) used to compute the logit-space velocity at the current time step. The angle head outputs seven torsional targets $a_i$ (three backbone angles plus up to four side-chain $\chi$ angles), with residue-specific masks disabling undefined torsions (e.g., glycine has no $\chi$ angles).

At each time step, these predictions define velocities on their respective manifolds: $\Omega_i$ for rotation, $t_i$ for translation, $\Delta \ell_i = \ell_{\text{pred}} - \ell_{\text{current}}$ for sequence, and wrapped angular differences for dihedrals. These velocities drive the ODE update during both training (for loss computation) and inference (for trajectory integration).

**Training objective and all-atom reconstruction.** We combine modality-specific conditional flow-matching losses with a stereochemical violation penalty inspired by OpenFold[25]:

$$\mathcal{L} = \lambda_\alpha \mathcal{L}_{\alpha_{\text{bb}}} + \lambda_{\text{SE(3)}} \mathcal{L}_g + \lambda_p \mathcal{L}_p + \lambda_\chi \mathcal{L}_{\alpha_\chi} + \mathcal{L}_{\text{viol}}.$$

where each modality loss measures the squared norm between predicted and analytic velocities in the appropriate tangent space. At each training step, we reconstruct atom37 coordinates from the current state $(\alpha_{bb}, g, p, \alpha_{sc})$ using standard protein geometry and calculate violation loss based on the reconstructed structure[25]. This term enforces physical plausibility during training by penalizing geometries that violate stereochemical constraints. We set hyperparameters as $\lambda_\alpha = 0.5$, $\lambda_{\text{SE(3)}} = 0.5$, $\lambda_p = 1.0$, $\lambda_\chi = 0.5$, and $\mathcal{L}_{\text{viol}} = 1$. All-atom reconstruction occurs mid-trajectory, enabling the model to learn how intermediate states affect final geometry rather than correcting violations only at t =1.

**Inference.** Sampling integrates learned velocities via Euler ODE steps with iterative geometric reconstruction. We discretize the time interval [0,1] into $K = 200$ uniform steps: $0 = t_0 < \cdots < t_K = 1$. At each step, we query the network for velocities $v_\theta(x_t, t, c)$ across all four modalities, then apply Euler updates:

Rotations: $R_{t+\Delta t} = \exp(\Delta t \Omega_t) R_t$

Translations: $t_{t+\Delta t} = t_t + \Delta t v_t^t$

Sequence logits: $\ell_{t+\Delta t} = \ell_t + \Delta t v_t^\ell$

Angles: $a_{t+\Delta t} = wrap(a_t + \Delta t v_t^a)$

After each step, we reconstruct atom37 coordinates from $(\alpha_{bb}, g, p, \alpha_{sc})$ to maintain geometric consistency. Deterministic integration eliminates the stochastic variance inherent to diffusion sampling.

**Datasets**. We assembled two receptor–peptide complex resources under a unified preprocessing pipeline: PepDB (binding-pocket-level complexes, n = 165 test targets) and PPDbench (whole-protein complexes, n = 613 test targets). Peptide sequences shorter than four residues were excluded; non-canonical residues were mapped to the nearest canonical amino acid. For each complex, we store residue amino-acid type, atom37 coordinates, backbone and side-chain torsion angles (as sin/cos pairs), chain and residue indices, and per-residue rigid-body frames derived from $C_\alpha$-C-N atom triplets.

PepDB provides pocket-localized receptor contexts and corresponding peptide binders, enabling training and evaluation of pocket-conditioned generators. To enable head-to-head comparison with PepFlow[19]. we adopt the identical train/validation/test partitioning protocol reported in that work, ensuring no target leakage. PPDbench comprises full receptor structures with peptide binders and supports whole-receptor-conditioned generation when binding-site annotations are unavailable[20]. We partition PPDbench at the complex level using an 8:1:1 random split (train:validation:test).

**Evaluation Metrics.** We assess designs via structural fidelity, interface quality, energetics, and distributional agreement. Structural metrics include $C_\alpha$ root-mean-square deviation (RMSD, Å) and TM-score (0-1 scale, higher indicates better global topology alignment). Interface metrics comprise binding-site ratio (BSR: fraction of designed residues within 5 Å of receptor) and buried surface area (BSA, Å). The secondary-structure similarity ratio (SSR) quantifies the proportion of matching secondary-structure elements between two peptides. Sequence metrics include amino-acid recovery (fraction matching native) and pairwise sequence diversity (mean Hamming distance within the 20-peptide ensemble per target).

For receptor-peptide energetics, we employ PyRosetta's InterfaceAnalyzer to compute binding free energy ΔG (Rosetta energy units, REU) and change in solvent-accessible surface area upon binding ΔSASA.

We use Kolmogorov–Smirnov (KS) two-sample tests to compare distributions of sampled dihedral angles against ground-truth distributions, reporting the maximum discrepancy D ∈ [0,1]. For joint $\varphi - \psi$ distributions (Ramachandran analysis), we compute Bhattacharyya coefficient (BC) and Jensen–Shannon similarity, both ranging [0,1] with 1 indicating perfect overlap. Per-target statistics are reported as mean ± SEM across 20 peptides per target, enabling comparison of central tendency and uncertainty.

## Results

**Performance under pocket versus whole-receptor conditioning.** We evaluated ApexGen on independent test with 20 peptides generated per target over 200 Euler ODE steps. We consider two conditioning regimes: pocket-conditioned (ApexGen-pocket, n = 165 targets) using localized binding-site contexts from PepDB, and whole-receptor-conditioned (APEXGen-protein, n = 613 targets) using full protein structures from PPDbench. Unless otherwise specified, statistics represent per-target means ± SEM.

Whole-receptor conditioning improves structural accuracy and interface quality while modestly reducing diversity (Table 1). ApexGen-protein reduces mean $C_\alpha$ RMSD from 3.58 Å to 3.20 Å (−10.6%), increases TM-score from 0.95 to 0.99 (+4.2%), and elevates BSR from 0.92 to 0.94 (+2.2%) compared to ApexGen -pocket. Stereochemical violation loss decreases 40.8% (0.106 → 0.063), indicating improved geometric plausibility. Sequence diversity declines modestly from 0.90 to 0.89 (−1.2%), reflecting tighter constraints from global receptor geometry (Figure 2a).

From a target-level success-rate perspective (Wilson 95% confidence intervals), the fraction of targets achieving RMSD ≤ 2 Å increases from 17.0% (ApexGen-pocket) to 31.5% (ApexGen-protein); RMSD ≤ 3 Å rises from 38.8% to 54.5%; and BSR ≥ 0.5 improves from 93.9% to 98.5%. TM-score ≥ 0.5 saturates at >99% for both regimes (Figure 2b,c). ApexGen-protein concentrates low-RMSD solutions in the quality–diversity space, whereas ApexGen-pocket explores more broadly, illustrating the classic quality–diversity trade-off (Figure 2d,e). The RMSD cumulative distribution function shows the ApexGen-protein curve dominates across all thresholds (Figure 2f), with systematically higher BSR (Figure 2g) and lower violation penalties (Figure 2h) under whole-receptor conditioning.

Comparing ApexGen-protein to PepFlow on the identical test split, we observe a striking divergence between local and global structural metrics (Table 1). PepFlow achieves slightly lower RMSD (3.03 Å versus 3.20 Å), suggesting better point-wise coordinate alignment. However, its TM-score of 0.33 indicates frequent topological failures, whereas ApexGen-protein maintains TM-score 0.99. This divergence suggests ApexGen prioritizes global fold topology and interface consistency (high TM-score, high BSR, low violation loss), whereas PepFlow may achieve local superposition quality at the cost of global structural agreement. The markedly higher interface metrics for ApexGen (SSR 0.99 versus 0.80, BSR 0.94 versus 0.84) further support this interpretation: ApexGen designs maintain tight receptor engagement throughout the peptide, while PepFlow designs exhibit more frequent interface dissociation. Full per-target comparisons appear in Supplementary Tables S1-S2.

**Backbone geometry fidelity.** We next examine how well the designed peptides reproduce native backbone torsion distributions. Under pocket conditioning, univariate Kolmogorov–Smirnov tests reveal close agreement for $\varphi$ and $\psi$ angles (D=0.086 and 0.093, respectively), whereas $\omega$ exhibits larger deviation (D=0.245; Figure 3a–c). Under whole-receptor conditioning, marginal discrepancies increase across all angles ($\varphi/\psi/\omega$: D = 0.21/0.13/0.32; Figure 3d–f), reflecting broader univariate distributions.

Analysis of joint $\varphi - \psi$ geometry using Ramachandran maps reveals the opposite trend. Normalized two-dimensional histograms show that whole-receptor conditioning achieves higher Bhattacharyya coefficient (0.668 versus 0.498) and Jensen–Shannon similarity (0.493 versus 0.387) compared to pocket conditioning (Figure 3g-j). Pocket conditioning enforces tighter marginal control over individual torsion angles but exhibits reduced overlap in the coupled $\alpha$-helix and $\beta$-sheet basins. Whole-receptor conditioning produces broader marginals yet better preserves the basin structure characteristic of native peptide conformations. These patterns suggest pocket conditioning applies stronger local regularization per angle, whereas whole-receptor conditioning captures global conformational preferences through richer geometric context, enabling the model to learn correlated backbone motions even with looser marginal constraints.

**Amino-acid composition.** ApexGen learns realistic amino-acid priors with context-linked, chemically grounded biases. For ApexGen-pocket, sampled amino-acid frequencies closely recapitulate the test-set distribution (Jensen–Shannon divergence JSD = 0.055; KL(sample‖GT) = 0.151), with residue-specific deviations concentrated in a few chemotypes (Figure 4a,b). The dominant shifts are depletion of Trp (−0.8 percentage points) and Cys (−0.5 pp) and enrichments in Glu (+1.2 pp), Gln (+0.9 pp), Ser (+0.8 pp), and Phe (+0.7 pp). These biases align with chemical liabilities: Cys introduces disulfide ambiguity absent explicit redox state annotations, and Trp's bulky indole side chain incurs steep stereochemical penalties in the violation loss. Conversely, enrichment of polar residues (Glu/Gln/Ser) and π-stacking Phe reflects a preference for hydrogen bonding and aromatic packing at peptide–receptor interfaces.

For ApexGen-protein, composition likewise remains close to ground truth (JSD = 0.039; KL(sample‖GT) ≈ 0.155) but shows systematic reweighting (Figure 4c,d): Thr (+1.1 pp), Asp (+0.9 pp), Leu (+0.8 pp), Phe (+0.7 pp), and Asn (+0.6 pp) are enriched, whereas Lys (−1.3 pp), Ser (−0.9 pp), Ile (−0.7 pp), Cys (−0.6 pp), Ala (−0.5 pp), and Val (−0.5 pp) are depleted. These shifts suggest a tilt toward polar, helix-permissive residues (Thr/Asp) and Leu-mediated hydrophobic packing, with under-sampling of basic Lys and β-branched aliphatics (Ile/Val). The persistent depletion of Cys and Trp across both conditioning regimes underscores a mechanistically interpretable bias rather than arbitrary drift. Collectively, both models preserve global amino-acid usage statistics (JSD < 0.06) while introducing plausible interface-driven adjustments.

**Boltz-2 structural predictions.** We assessed the structural plausibility and predicted binding propensity of designed complexes using Boltz-2[26], a deep-learning model for protein–peptide structure and affinity prediction. Across all six evaluation metrics, designed peptides match or exceed the distributions of native binders (Figure 5). Predicted binding affinity distributions are nearly identical (designed mean −0.443 versus native −0.386), with both peaking near −1, consistent with moderate binding strength characteristic of transient peptide–protein interactions. Designed peptides show slight enrichment in binding probability (fraction with P_bind > 0.5 increases 8%), indicating Boltz-2 assigns higher confidence to productive binding modes.

Structural confidence metrics overlap extensively: complex pLDDT distributions are indistinguishable (designed mean 78.3 versus native 77.9), as are overall confidence scores (designed 0.71 versus native 0.70). The predicted TM-score (pTM) for peptide backbones is nearly identical between designed and native sets (both ≈ 0.68), underscoring geometric realism of generated folds. The interface predicted TM-score (ipTM) distributions exhibit mild enrichment at intermediate-high values for designs (mode 0.55 versus 0.50 for native), reflecting productive interfacial packing recognized by Boltz-2's learned interface geometry patterns.

Together, these results demonstrate that ApexGen generates peptide binders with binding propensities, structural integrity, and model confidence levels comparable to natural complexes. However, we caution that Boltz-2 is trained on PDB-derived data overlapping with our training distribution, potentially introducing circularity: high scores may reflect "native-like appearance" to a model with shared training priors rather than true binding affinity. We interpret these metrics as evidence of geometric plausibility and stereochemical realism, complementing but not replacing experimental validation.

**Rosetta binding energetics.** We assessed designed complexes using PyRosetta's InterfaceAnalyzer, computing binding free energy (ΔG, Rosetta energy units) and buried solvent-accessible surface area (ΔSASA). Across both conditioning regimes, ΔG distributions for designed complexes are broader than ground truth but shifted toward more favorable energies (Figure 6a,c). Mean energies are significantly lower for designs under one-sided Welch t-tests (ApexGen-protein: $p = 1.1 \times 10^{-3}$; ApexGen-pocket: $p = 1.7 \times 10^{-11}$), indicating ApexGen samples populate low-energy basins more frequently than native complexes under identical Rosetta scoring.

Interface burial is reduced and more narrowly distributed for designs (left-shifted ΔSASA in both pocket and protein definitions; Figure 6b,d). At the whole-protein level, the designed distribution peaks near 1000, typical of compact peptide–protein interfaces. In the pocket-restricted calculation (considering only pocket-proximal residues), ΔSASA is systematically smaller, but the same left-shifted trend persists.

We interpret these findings as evidence that ApexGen generates peptide binders with more favorable Rosetta energies while maintaining physically plausible, compact interfaces, consistent with tightly packed contacts that achieve favorable ΔG without inflating buried surface area. We also interpret Rosetta scores as measures of geometric plausibility and stereochemical realism within the learned energy landscape, not as absolute affinity predictions requiring experimental validation.

## Discussion

ApexGen introduces deterministic, manifold-aware co-design of peptide sequence and structure via receptor-conditioned flow matching. By learning four coupled vector fields across backbone torsions, rigid-body frames, amino-acid identities, and side-chain angles, with geometric plausibility enforced through mid-trajectory all-atom reconstruction, ApexGen overcomes three key limitations of prior methods: it avoids decoupling sequence and structure, eliminates stochastic sampling noise, and ensures chemical validity without post hoc correction. As a result, ApexGen designs achieve near-native folds (mean TM-score 0.99 versus 0.33 for baselines such as PepFlow) and strong receptor engagement (BSR 0.94) while sampling deterministically via integration of a learned ODE.

Using the full receptor structure (ApexGen-protein) substantially improves results over pocket-only conditioning. The RMSD≤2 Å success rate jumps from 17.0% to 31.5%, and stereochemical violations drop by 40.8%, showing that global context provides stronger guidance. Interestingly, the two regimes enforce backbone geometry differently: pocket conditioning yields tighter marginal angle distributions (lower KS scores for $\varphi/\psi$) but poorer joint $\varphi - \psi$ overlap (Bhattacharyya coefficient (BC): 0.498), whereas whole-receptor conditioning yields broader marginals (D = 0.21/0.13) but better preservation of native Ramachandran basins (BC: 0.668). This suggests pocket conditioning applies strong local regularization on each dihedral, while whole-protein conditioning captures correlated backbone motions through the richer structural context. The amino-acid biases further illustrate chemically interpretable learning: ApeGen avoids bulky Cys/Trp (hence their depletion) and favors polar/aromatic residues for interface bonding. These are not arbitrary drifts but reflect our model's generic stereochemical penalties and the physics of interfaces.

Compared to PepFlow, ApexGen's integrated trajectory pays off in topology. PepFlow achieves marginally lower RMSD (3.03 Å vs 3.20 Å) but catastrophically low TM-scores (0.33), indicating wrong folds. Its decoupled backbone–sequence design appears to sacrifice global consistency. In contrast, ApexGen maintains sequence–structure coupling throughout training. The simultaneous updates and stereochemical penalties propagate corrections across modalities, leading to globally coherent designs. The deterministic integration also removes sampling variance, removing the need for a separate sequence model (such as ProteinMPNN) and simplifying the pipeline.

Nonetheless, some biases of ApexGen may limit certain applications. In particular, the strong learned depletion of Cys (–0.5 to –0.6 points) and Trp (–0.8 points) could hinder designs that truly require these residues (e.g. disulfide-stapled peptides or tryptophan anchors). This arises because our training objective penalizes stereochemical strain without explicit disulfide or packing constraints. Remedies could include adding disulfide constraints in training (treating Cys pairs as fixed edges), reweighting these residues during inference, or grafting them post-design via repacking. For many peptide therapeutic targets (helix-mediated PPI inhibitors, etc.) the standard amino-acid set is sufficient, but expanding to noncanonical residues will be important in future work.

Finally, the Rosetta and Boltz-2 evaluations suggest ApexGen designs have favorable energetics and native-like confidence, but we emphasize these as indicators of quality, not definitive proof of binding. Boltz-2's training overlap may inflate apparent affinity, and Rosetta scores depend on the force field. Ultimately, empirical validation (e.g. SPR binding assays, crystallography) is needed. In silico, these analyses show ApexGen designs occupy plausible low-energy, well-packed states appropriate for experimental follow-up.

In summary, ApexGen addresses the computational bottleneck in peptide drug discovery by rapidly generating receptor-specific peptide candidates with explicit 3D structures. It complements experimental approaches: unlike sequence libraries (phage/mRNA display), it provides immediate structural hypotheses; unlike pure docking or MD, it covers vast sequence space efficiently. Moving forward, ApexGen could be integrated into multi-stage pipelines: for example, using pocket conditioning to generate diverse scaffolds, followed by whole-protein conditioning for refinement. Future extensions may incorporate explicit affinity or specificity objectives, expand the residue alphabet (cyclic peptides, D-amino acids), and adopt active learning loops with experiments. With continued development and validation, ApexGen could significantly accelerate the design of peptide therapeutics targeting challenging protein–protein interfaces.

## Conclusion

ApexGen provides a deterministic, geometry-aware approach to peptide sequence-structure co-design. By unifying backbone and side-chain generation in a receptor-conditioned flow matching framework, it achieves near-native peptide designs with high interface complementarity. This capability could expedite early-stage discovery of peptide drugs by producing diverse, structure-informed candidates for downstream testing. As with any computational design, empirical validation and medicinal chemistry optimization are essential next steps. Nonetheless, ApexGen demonstrates that combining modern generative modeling with manifold geometry can greatly expand the accessible peptide design space, potentially unlocking new therapeutics for previously intractable targets.

## Acknowledgements

Cesar de la Fuente-Nunez holds a Presidential Professorship at the University of Pennsylvania. Research reported in this publication was supported by the NIH R35GM138201 and DTRA HDTRA1-21-1-0014. We thank de la Fuente Lab members for insightful discussions.

## Conflict of interest

Cesar de la Fuente-Nunez is a co-founder of, and scientific advisor, to Peptaris, Inc., provides consulting services to Invaio Sciences, and is a member of the Scientific Advisory Boards of Nowture S.L., Peptidus, European Biotech Venture Builder, the Peptide Drug Hunting Consortium (PDHC), ePhective Therapeutics, Inc., and Phare Bio.

# Figures

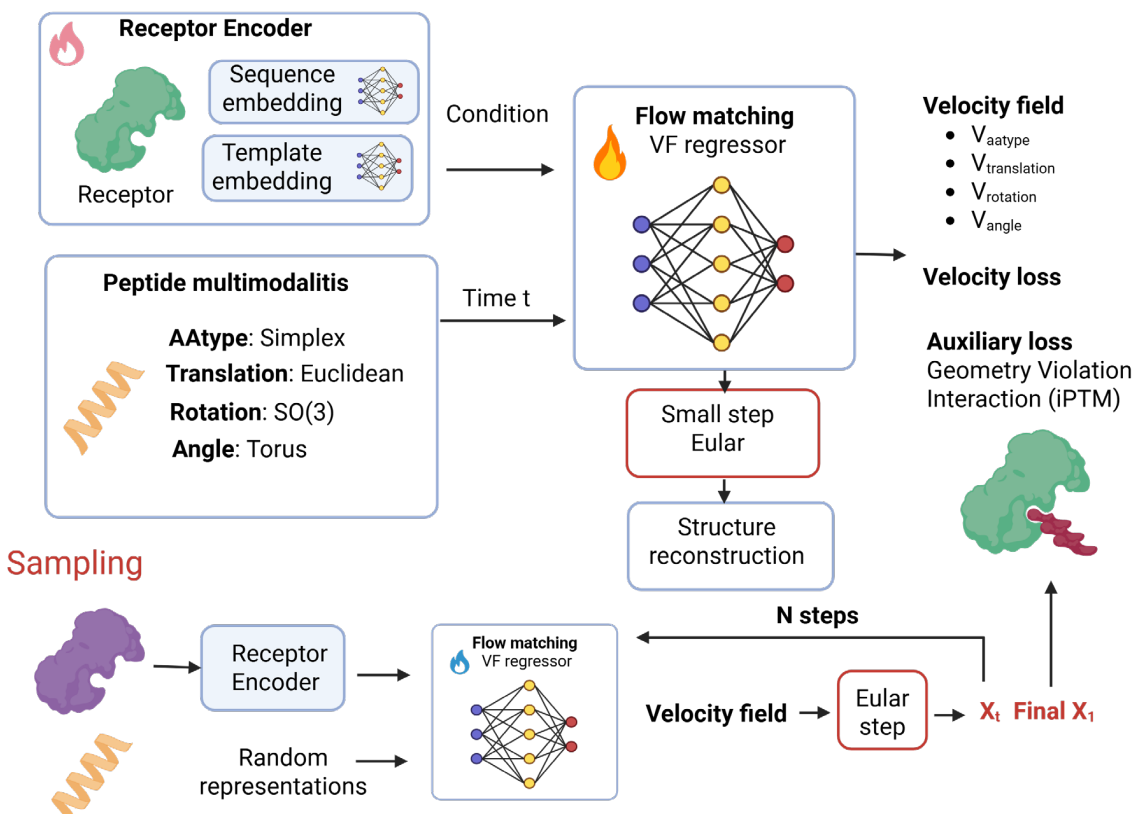

**Figure 1. ApexGen training and sampling workflow for receptor-conditioned peptide design.**
ApexGen training and sampling workflow for receptor-conditioned peptide design. (Top) During training, the receptor protein structure conditions a flow-matching network that predicts velocity fields for four peptide modalities (sequence, translation, rotation, dihedral angles). These predicted velocities are trained against the analytic conditional flow-matching targets, with an auxiliary stereochemical violation penalty computed from reconstructed all-atom coordinates. (Bottom) During sampling, random initial peptide states are iteratively refined through $N$ Euler ODE integration steps using the learned velocity field, yielding final designed peptide-protein complexes.

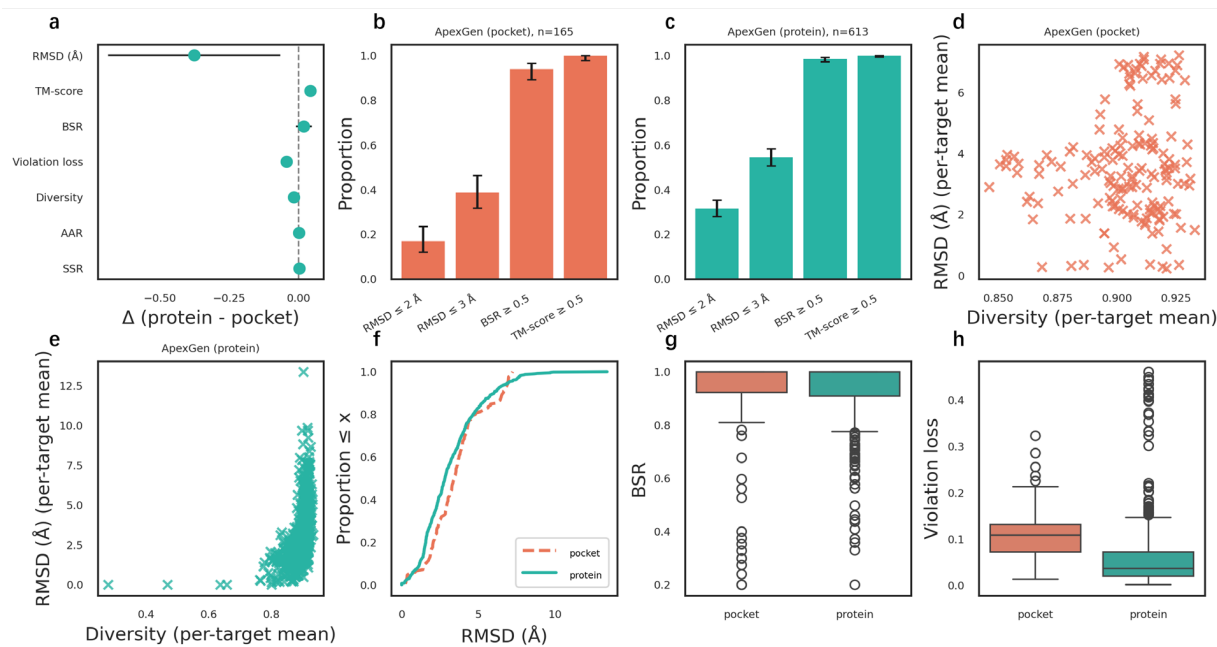

**Figure 2. Impact of conditioning context on design performance. (a)** Mean difference (ApexGen-protein minus ApexGen-pocket) for each metric across targets (error bars: 95% CI). **(b,c)** Target-level success rates under pocket (b, n=165) and whole-protein (c, n=613) conditioning, for RMSD thresholds and binding-site ratio (BSR) thresholds; error bars are Wilson 95% CI. **(d,e)** Quality–diversity plots (per-target best-of-20 RMSD vs. mean sequence diversity) for pocket **(d)** and whole-protein **(e)** designs. **(f)** Cumulative distribution of per-target best RMSD: the whole-protein (teal) curve dominates pocket (salmon) at all thresholds. **(g,h)** For each target, 20 peptide binders were generated using our model, and the mean value of each metric was computed. The boxplots show the distribution of these per-target averages across all targets: **(g)** binding success rate (BSR) and (h) violation loss. Whole-protein conditioning consistently achieves higher BSR and lower violation loss compared to pocket conditioning.

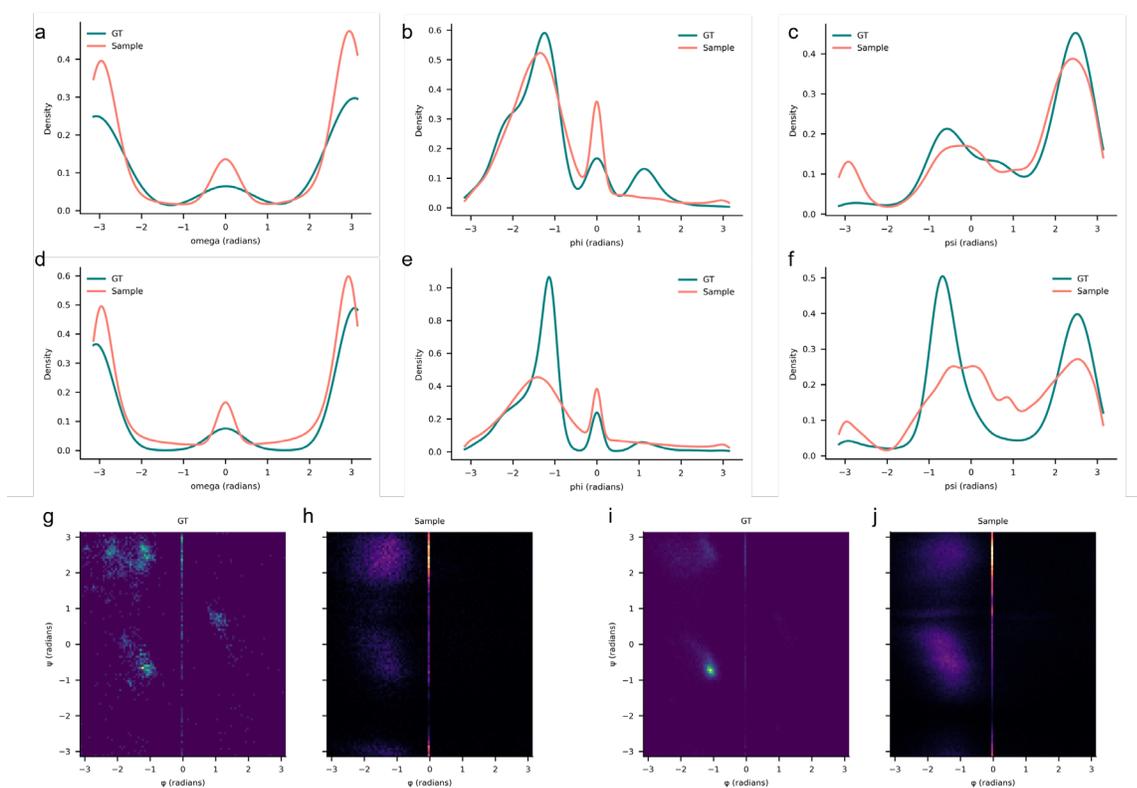

**Figure 3. Backbone torsion-angle fidelity under pocket versus protein conditioning.** **(a–c)** Distributions of $\omega$, $\varphi$, and $\psi$ angles for native peptides (teal) and pocket-conditioned designs (salmon). **(d–f)** Same for whole-protein designs (teal) vs. native. **(g, h)** Ramachandran $\varphi - \psi$ plots: ground truth **(g)** and pocket-conditioned designs **(h)**. **(i,j)** Ground truth **(i)** and whole-protein designs **(j)**. Pocket designs have tight marginal angle distributions but reduced overlap of $\alpha$-helix and $\beta$-sheet regions. Whole-protein designs have broader marginals with stronger overlap in the native basins. Overall, pocket conditioning enforces stricter local $\varphi/\psi$ agreement, while whole-protein conditioning better preserves the joint basin structure.

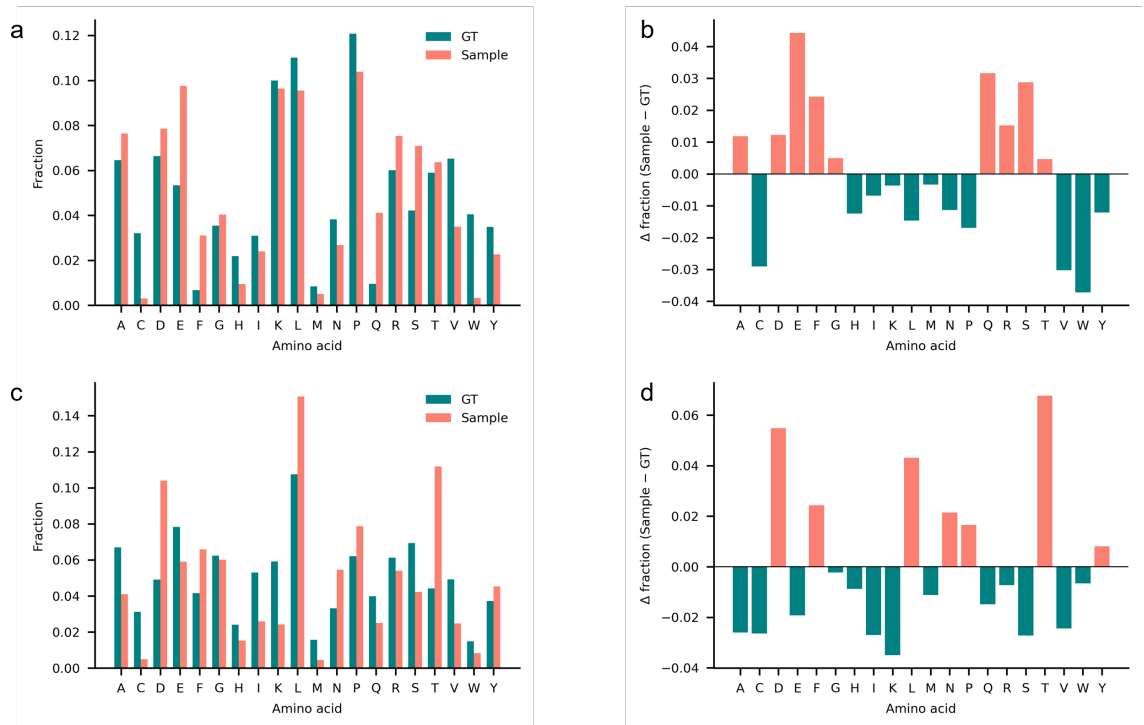

**Figure 4. Amino-acid composition of designed peptides. (a)** Per-residue frequency of each amino acid for native peptides (GT, teal) and pocket-conditioned designs (Sample, salmon). **(b)** Difference (Sample-GT) for pocket conditioning. **(c, d)** Same plots for whole-protein conditioning. Positive values indicate residues enriched in the designs. Both regimes preserve overall distributions (low JS divergence) but show depletion of Cys/Trp and enrichment of polar/aromatic residues (see text).

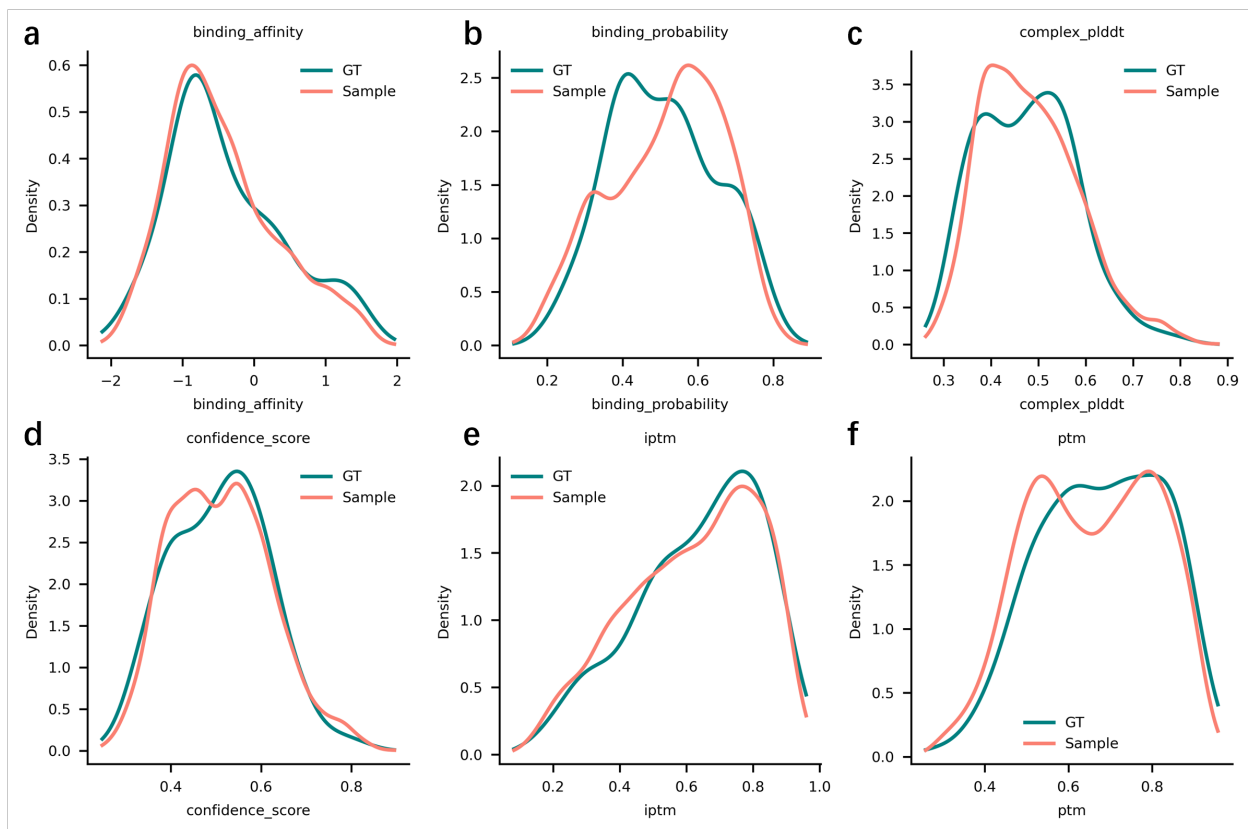

**Figure 5. Boltz-2 evaluation of designed peptide–protein complexes.** Kernel density estimates compare designed peptides (Sample, salmon) with native complexes (GT, teal). **(a)** Predicted binding affinity (Boltz-2 score); **(b)** binding probability $P_{bind}$; **(c)** complex pLDDT; **(d)** overall confidence score; **(e)** interface predicted TM-score (ipTM); **(f)** peptide backbone predicted TM-score (pTM). The designed and native distributions are largely overlapping, indicating similar predicted binding strength and model confidence. Designs show a slight enrichment for higher $P_{bind}$ and ipTM.

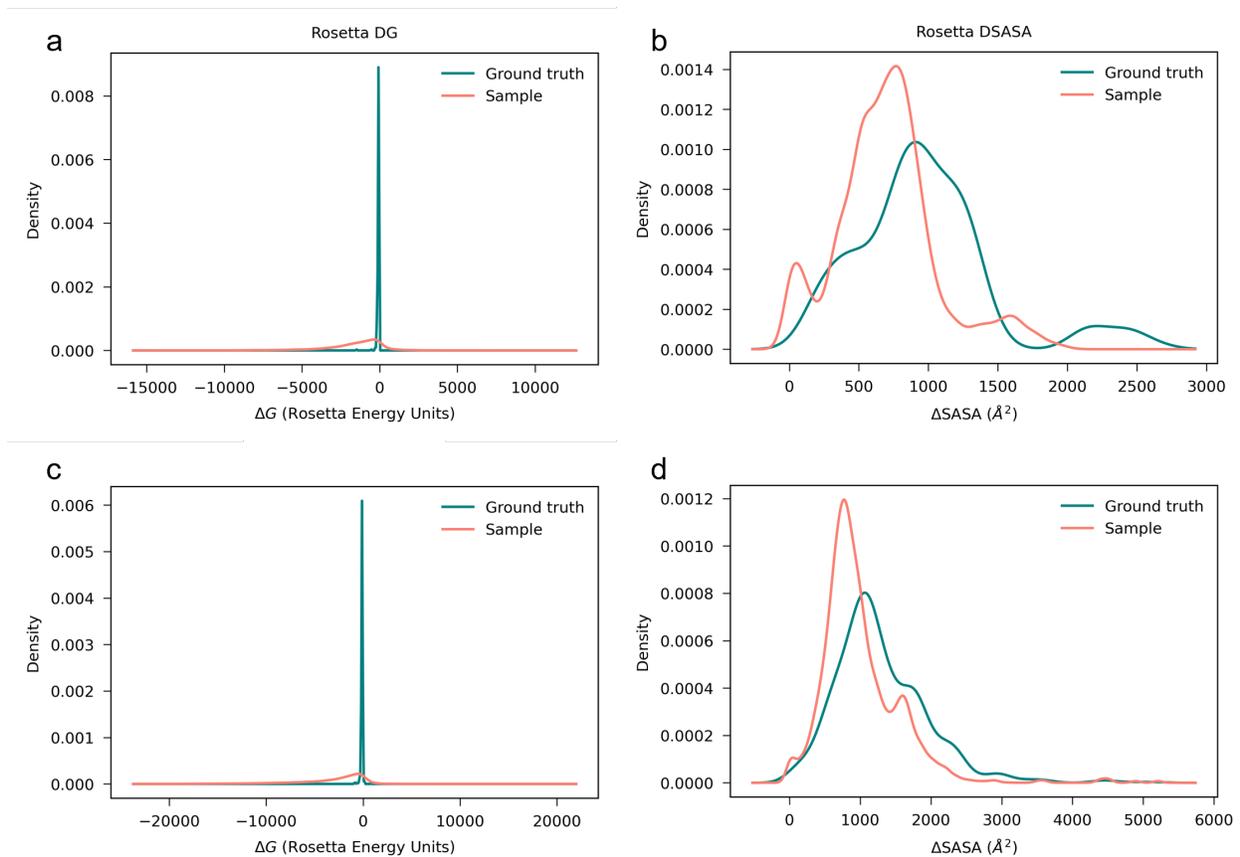

**Figure 6. PyRosetta InterfaceAnalyzer evaluation.** Kernel density plots compare native (GT, teal) and designed (Sample, salmon) complexes under Rosetta scoring. **(a)** Pocket-conditioned model: binding free energy ΔG (REU): Designed peptides show a broader distribution shifted toward lower (more favorable) energies. **(b)** Pocket-conditioned model: buried surface area ΔSASA (Å²). Designs exhibit a slightly reduced interface area. **(c)** Protein-conditioned model: binding free energy ΔG. The low-energy shift for designs persists. **(d)** Protein-conditioned model: buried surface area ΔSASA. Designs display moderately smaller burial. Overall, the designed complexes tend to be more compact and energetically favorable according to the Rosetta scoring model.

# Tables

**Table 1. Comparative performance of ApexGen (pocket) and ApexGen (protein) versus PepFlow on the independent test set**. Values are means over targets (20 peptides each). SSR: secondary-structure similarity ration; BSR: binding-site ratio; RMSD (Å); Diversity: mean pairwise sequence diversity; TM-score; AAR: amino-acid recovery; Violation loss: stereochemical penalty (lower is better). Sample sizes: n=165 (pocket), n=613 (whole).

| Metric | ApexGen (pocket) | ApexGen(protein) | PepFlow |
| --- | --- | --- | --- |
| SSR | 0.9824 | 0.9857 | 0.7950 |
| BSR | 0.9201 | 0.9391 | 0.8377 |
| RMSD (Å) | 3.5757 | 3.1988 | 3.0344 |
| Diversity | 0.9037 | 0.8869 | 0.7428 |
| TM-score | 0.9513 | 0.9937 | 0.3260 |
| AAR | 0.0708 | 0.0725 | 0.1190 |
| Clashes | 0 | 0 | – |
| Violation loss | 0.1061 | 0.0628 | – |
| GT violation loss | 0.3895 | 1.1514 | – |